\documentclass[]{spie}

\usepackage[utf8]{inputenc}
\usepackage[numbers]{natbib}
\usepackage{setspace}
\usepackage{makecell}
\usepackage{nicefrac}
\usepackage{graphicx}
\usepackage{url}
\usepackage{cleveref}
\usepackage{subcaption}

\usepackage{enumitem}
\setlist{nosep}

\title{RF-ICE: large-scale gigahertz readout of frequency-multiplexed microwave kinetic inductance detectors}
\author[a]{M. Rouble}
\author[b]{G. Smecher}
\author[c]{A. Anderson}
\author[d,e]{P. S. Barry}
\author[f]{K. Dibert}
\author[a, g]{M. Dobbs}
\author[f]{K. S. Karkare}
\author[a]{J. Montgomery}

\affil[a]{McGill University, Canada}
\affil[b]{Three-Speed Logic, Inc., Canada}
\affil[c]{Fermi National Accelerator Laboratory, United States}
\affil[d]{Argonne National Laboratory, United States}
\affil[e]{Cardiff University, United Kingdom}
\affil[f]{The University of Chicago, United States}
\affil[g]{Canadian Institute for Advanced Research, Program in Gravity and the Extreme Universe, Canada}

\date{June 2022}

\begin{document}

\maketitle
\begin{abstract}
    We present RF-ICE, a novel readout platform for microwave kinetic inductance detectors (MKIDs), optimized for use on millimeter-wavelength telescopes. The RF-ICE system extends ICE, a versatile, mature signal processing platform currently in use on telescopes around the world, into a new operational domain with MKIDs biased with gigahertz carriers. The system couples the FPGA-based ICE motherboard with a radio-frequency digitization daughterboard to enable direct digital synthesis from 0 to 6 GHz without the need for external mixing. The system operates two independent readout modules, each with 1024 frequency-multiplexed readout channels spaced across 500 MHz of carrier bandwidth. The system, which is under active development, is in operation with prototype detector wafers and will be deployed for the upcoming SPT-SLIM and SPT-3G+ experiments.
\end{abstract}

\newpage

\section{Introduction}\label{sec:intro}

The ICE readout framework is a versatile, mature signal processing platform currently in active use on telescopes around the world, observing at radio and millimetre wavelengths \cite{Bandura2016}. It consists of an field-programmable gate array (FPGA) motherboard (the ``IceBoard''), which pairs with a daughterboard via a standard FPGA-mezzanine connector (FMC) to adapt to each use case. RF-ICE extends the foundational hardware, software, and firmware of the ICE system to a new operational domain at gigahertz frequencies, operating large arrays of microwave kinetic inductance detectors (MKIDs) as part of the upcoming generation of millimetre and sub-millimetre cosmology experiments on the South Pole Telescope.

The South Pole Telescope (SPT) is a 10m telescope located at the Amundsen-Scott South Pole Station \cite{spt2011}. Since its commissioning in 2007, it has hosted three primary experiments: SPT-SZ \cite{sptsz}, SPTpol \cite{sptpol}, and (presently) SPT-3G \cite{spt3g}, mapping the cosmic microwave background (CMB). As the sensitivity of modern CMB experiments is photon-statistics-limited, rather than by the instrument or individual detectors, improvements in sensitivity are to be made by increasing the number of detectors operated by an experiment. Using frequency-multiplexed readout, each generation of experiment has increased its detector count over its predecessor, culminating in the more than 16 000 transition edge sensors (TESes) on the SPT-3G focal plane today \cite{spt3g}. Future experiments will continue to seek higher sensitivity through further increases in focal plane density. This motivates the use of a detector technology which can be readily and reliably fabricated at scale, to avoid the cost and time associated with the complex lithography of TESes and to accelerate the prototyping and testing process. 

The next generation of experiments to be operated on the SPT will be the South Pole Telescope Summertime Line Intensity Mapper (SPT-SLIM; \cite{karkare2021}), a pathfinder for on-chip spectrometers, and SPT-3G+, which will replace the current SPT-3G survey with a new receiver targeting precision measurements of the kinetic Sunyaev-Zel'dovich effect and a first detection of Rayleigh scattering of the CMB \cite{Anderson2022}. These new cameras will use MKIDs exclusively.

MKIDs are emerging as a promising superconducting detector for cosmological and astronomical observations, though their large-scale deployment on the sky at microwave frequencies is in a nascent stage. Consisting of a light-sensitive superconducting resonator, MKIDs require no additional cryogenic multiplexing hardware and can be read out via a traditional transistor-based low-noise amplifier, offering a drastically simplified cryogenic readout architecture in comparison with TES-based systems, which require separate filter modules and/or amplification through superconducting quantum interference devices. Cheaper and more straightforward to fabricate than TESes, MKIDs can also achieve significantly higher mapping speeds through higher multiplexing factors and higher focal plane densities at frequencies above 200 GHz, and offer a rapid turnaround time between prototype design, fabrication, and testing. The mechanism of detection for MKIDs is through the monitoring of the location of the resonant feature associated with each device. Energy from incident photons alters the kinetic inductance of the detector, changing this resonant frequency. A readout system therefore measures the phase and/or amplitude of a probe tone placed near the resonance to determine the optical power on the device.

The development of these upcoming experiments is active and ongoing, as is the development of their readout systems. Informed by the design choices and specifications of the upcoming SPT-SLIM and SPT-3G+ experiments, as well as by those that have come before them, RF-ICE builds on the foundational ICE readout framework to meet the challenges of this new domain.

\section{RF-ICE}\label{sec:rf-ice}

\subsection{Overview}

\begin{figure}
    \centering
    \includegraphics[width=0.8\linewidth]{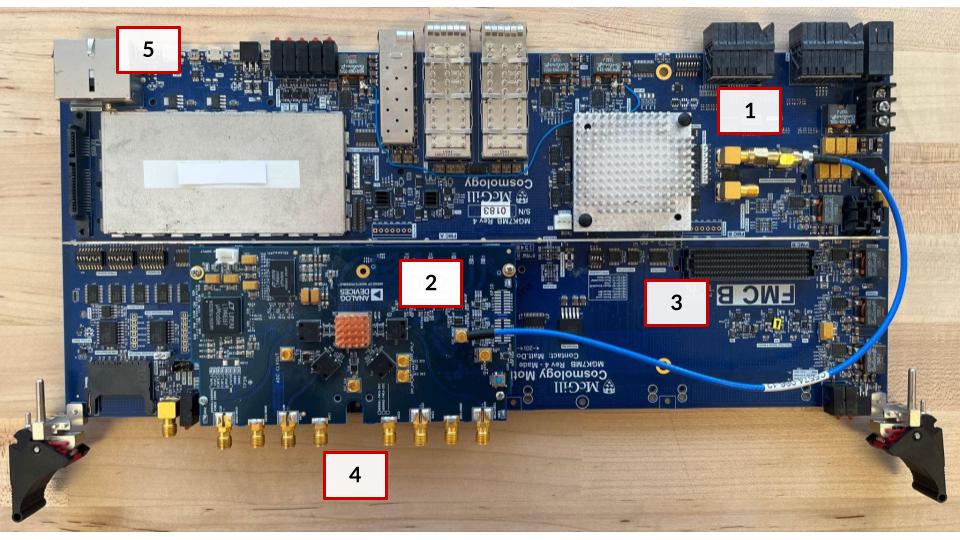}
    \caption{The digital hardware of the RF-ICE platform: an ICE motherboard (1; the ``IceBoard") with Analog Devices AD9082-FMCA-EBZ mezzanine (2). A multi-purpose signal processing platform, the IceBoard pairs with the mezzanine through one of its two available standard FMC interfaces, one of which is visible as (3). The mezzanine is an evaluation board for the AD9082 mixed-signal front-end device, housing two independent readout modules which interface with external analog readout electronics via SMA connectors (4). The IceBoard communicates with a control PC via wired Ethernet (5).}
    \label{fig:iceboard}
\end{figure}

RF-ICE pairs the ICE motherboard \cite{Bandura2016}, a signal processing platform centred on the Xilinx Kintex-7 FPGA, with the Analog Devices AD9082-FMCA-EBZ mezzanine card \cite{ad9082board}. The mezzanine is an evaluation board for the AD9082 \cite{ad9082}, an integrated mixed signal front-end device comprising four 16-bit 12 GSPS digital-to-analog converters (DACs) and two 12-bit 6 GSPS analog-to-digital converters (ADCs). The RF-ICE system makes use of the ICE platform's versatility, mating the commercially-available mezzanine to the IceBoard through its standard FPGA-mezzanine connector (FMC) interface and allowing rapid progression from conception to implementation without fabricating additional custom hardware. An ICE motherboard with AD9082 mezzanine is pictured, on the desktop, in Figure \ref{fig:iceboard}.

Each IceBoard is paired with one AD9082 mezzanine, which houses two ``readout modules'', independent signal processing chains operating from 0 to 6 GHz, each simultaneously synthesizing up to 1024 channels in firmware over an instantaneous bandwidth of 500 MHz to give a total possible detector count per motherboard of 2048. Each module comprises two DACs (a ``carrier'' and a ``nuller'') and one ADC, with a numerically-controlled oscillator (NCO) which determines the placement of the instantaneous bandwidth. Although not in use at present, the nuller synthesis pathways allow the use of active feedback for nulling carrier tones to improve amplifier stability, or to implement continuous tone tracking. The system uses direct sampling for synthesis and demodulation, without the need for external quadrature mixing.

Designed to be mounted in a ``crate'' (a high-density equipment rack housing an array of IceBoards), the system receives its 18V power supply either through a custom crate backplane or through attachment points on the motherboard, and communicates with a control computer over a wired Ethernet connection. A single, experiment-wide clock prevents intermodulation products from, for example, peripherals or beating oscillators, and can be sourced either from the motherboard's onboard precision oscillator or from an external source. Each readout module interfaces with the analog electronics of the external signal chain through SMA connectors on the mezzanine.

\subsection{Firmware and digital signal path}

\begin{figure}
    \centering
    \includegraphics[width=\textwidth]{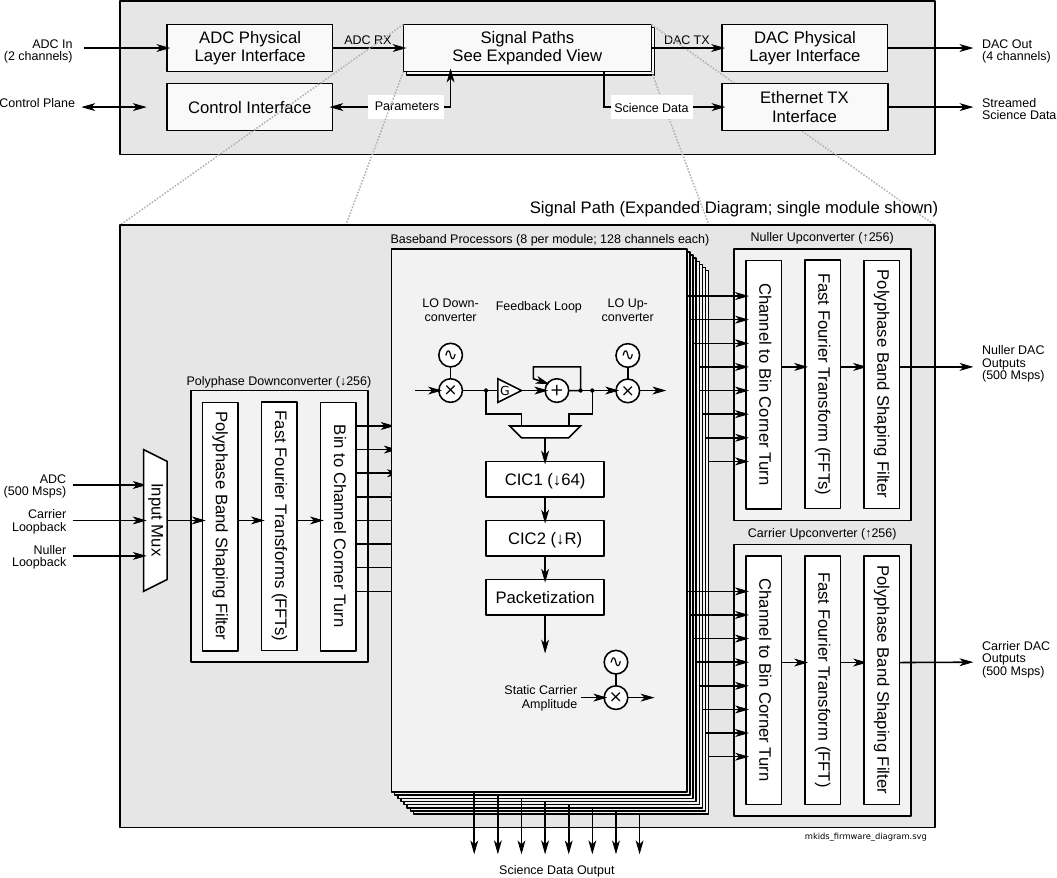}
    \caption{
        Signal path for a single RF-ICE comb. There are two such signal paths per IceBoard.
        The FPGA receives 16-bit I/Q data at 500 MSPS from the ADC via a JESD204B link.
        This datastream has already been downconverted from a sampling frequency of 6 GSPS on the ADC; hence, MKIDs resonators may be placed anywhere in a contiguous 500 MHz of RF spectrum (including higher Nyquist zones) subject to the performance of the ADC/DAC and surrounding analog chain.
        Data is channelized via a $2 \times$ oversampled polyphase filterbank (PFB)\cite{Harris2021}, producing 512 subbands at approximately 2 MSPS each.
        Due to the design of the PFB and window function, any frequency across the 500 MHz input spectrum appears in 2 subbands.
        For a given detector, the subband with the highest signal amplitude at the detector frequency is chosen and downconverted using a complex DDS/mixer.
        Up to 1024 complex baseband channels are supported per multiplexed comb of MKIDs detectors.
        An optional nuller loop may be used to linearize amplification stages (such as HEMTs).
        Readout data is sampled from each baseband channel at a variable decimation rate between 64 (R=1) and 4096 (R=64).
        Carrier and nulller synthesis follows the same path in reverse: baseband IQ samples at 2 MSPS are converted to subbands using a DDS upconverter; subbands are combined into a 500 MSPS complex datastream using an oversampled PFB.
        The resulting 500 MSPS I/Q samples are transmitted to DACs using JESD204B links.
        Finally, they are upconverted to 6 GSPS on the DACs and passed to the system's analog outputs.
    }
    \label{fig:dan-path}
\end{figure}

The RF-ICE readout firmware is shown in \Cref{fig:dan-path}.
This firmware is chiefly differentiated by the use of polyphase filter-bank (PFB) converters \cite{Harris2021} on both demodulation and synthesis paths.
PFB demodulators are in conventional among frequency-domain multiplexed TES \cite{Smecher2022b} and MKID  \cite{Gordon2016, Duan2010} readout.
However, PFBs in MKIDs synthesizer paths are relatively unusual, and provide two key benefits:
\begin{itemize}
    \item Synthesizer parameters may be continuously varied in real time, allowing feedback-control schemes such as continuous tone tracking and dynamic nulling, and
    \item By using a highly-decimated timestream representation for each channel, PFB synthesizers eliminate bandwidth and data-storage bottlenecks associated with static playback buffers for bias tones.
\end{itemize}

The AD9082 analog front-end communicates with the FPGA via JESD204B\cite{jesd204b} links (8 lanes; 10 GSPS per lane).
RF signals are digitized by the ADC at 6 GSPS, mixed down with an on-chip digital LO, and decimated on the transceiver to 500 MSPS IQ samples.
Each IceBoard receives 2 such ADC channels from a single RF front-end, each of which encodes an MKIDs comb at a multiplexing factor of up to 1024. The synthesis path follows the same process in reverse; 500 MSPS IQ samples from the FPGA are transferred via JESD204B to the AD9082, where they are upconverted to 6 GSPS, mixed up with the integrated digital LO, and emitted by the DAC.

Within the FPGA, each 500 MSPS data stream is demodulated to 1024 baseband channels using a 2-stage process:
\begin{itemize}
    \item First, a $2 \times$ oversampled PFB \cite{Harris2021} converts the 500 MSPS timestream into 512 subbands, centered at multiples of 976.5625 kHz.
    \item Second, a given frequency is converted to baseband by selecting the nearest subband centre and applying a locally generated DDS carrier to complete the downconversion.
\end{itemize}

Baseband channels are sampled at $500 \times 10^6 \div 256 \approx 2~\mathrm{MSPS}$.
To allow a resource-efficient implementation of baseband algorithms, channels are processed in 128-channel blocks using an internal DSP clock of 250 MHz.
To achieve a total multiplexing factor of 1024, there are 8 such baseband blocks per MKIDs comb.

Science data is generated by further decimation of baseband signals using a chain of 2 Cascaded Integrator-Comb (CIC) filters.
The first (CIC1) is a fixed-rate, 64-times decimator.
The second (CIC2) allows a variable decimation rate of $R=\{1, 2, 4, 8, 16, 32, 64\}$.
Each baseband block packetizes a separate data stream from CIC2, which includes an IRIG-B timestamp\cite{irigb}.
The resulting packets are combined and streamed across a 1 Gb/s Ethernet link.

When integrated into a full readout system, compensation for passband droop intrinsic to CIC filters is necessary.
In the ICE firmware implementation for TES readout\cite{Smecher2022b}, such filters were previously integrated into the FPGA and part of the signal chain in \Cref{fig:dan-path}.
These FIR filters are not present in the RF-ICE firmware design due to the increased design pressure on the FPGA's block RAMs.
Because an additional FIR decimator is already part of the external readout chain, this design assumes CIC compensation can be rolled into the existing (external) FIR without difficulty.
Because the computational throughput is not high, such an FIR may alternatively be implemented on FPGA-adjacent processors (such as the ICE system's separate ARM system-on-chip, or the integrated ARM cores in more recent FPGA System-on-Chips.)

\begin{figure}
    \centering
    \includegraphics[width=0.5\textwidth,angle=90]{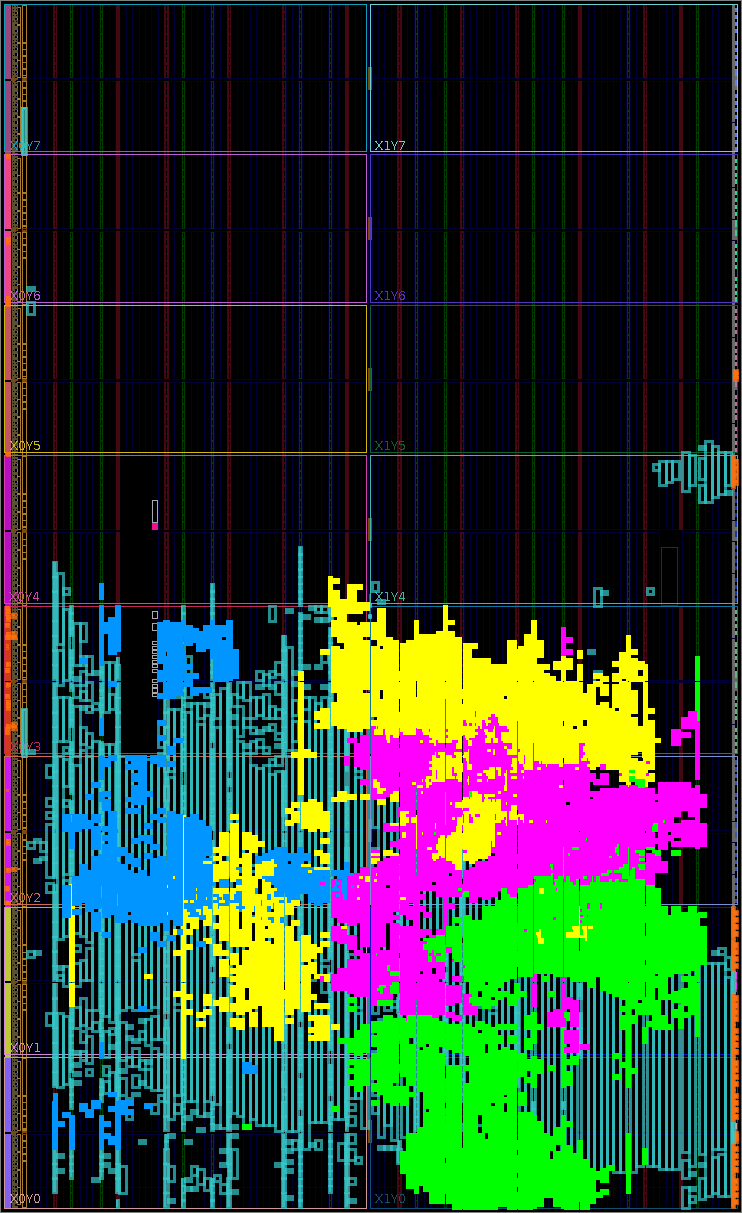}
    \caption{
        Top-level floorplan of the Kintex-7 '420T FPGA, showing a single RF-ICE readout comb at 1024x multiplexing factor.
        Unused FPGA resources are dark (50\% of the FPGA is reserved for the second comb); resources that are used but not explicitly highlighted are in teal.
        PFB up-/down-converters have been highlighted to show relative resource usage:
        the coarse downconversion is coloured green;
        the nuller coarse upconverter is coloured yellow; and
        the carrier coarse upconverter is coloured pink.
        One of the eight baseband blocks is highlighted in blue; although the placement for these 8 blocks varies, their resource utilization is the same.
    }
    \label{fig:floorplan-top}
\end{figure}

A floorplan of the IceBoard's FPGA configured for a single RF-ICE readout module at a multiplexing factor of 1024x is shown in \Cref{fig:floorplan-top}.
Striping evident in floorplan images show the design's appetite for both DSPs (narrower vertical stripes) and block RAMs (wider vertical stripes) compared to ordinary fabric, which tends to form less differentiated structures.

Top-level resource usage for a 1-comb, 1024x RF-ICE build is shown in \Cref{tab:resource-usage}.
This design relies heavily on both DSP and BRAM elements within the FPGA fabric.
Scaling to higher multiplexing factors (via additional baseband blocks) or additional resonator combs (via replicating the entire signal path) would require more of both DSPs and BRAMs on a larger FPGA.
The FPGA used in the IceBoard is the Xilinx Kintex-7 '420T, which was a midrange FPGA when it was released in 2011; modern midrange devices are many times more capable.

Resource consumption is also highly dependent on the bit widths of signals across the datapath, which primarily affects the efficiency with which the FPGA's DSPs and BRAMs are utilized. Because the natural port width of DSPs does not neatly match the natural port width of BRAMs, picking an optimal bit width presents a challenge. Where possible, the signal path uses 24-bit widths for each of the I and Q components for all baseband signals at 2 MSPS.
DAC and ADC signal paths are 16 bits wide.

\begin{table}
	\centering
	\begin{tabular}{l|l|l|l}
		\hline
		\textbf{Resource} & \textbf{Utilization} & \textbf{Available} & \textbf{\% Used} \\ \hline
		Slice LUTs & 62,993 & 260,600 & 24\% \\
		Slice Registers & 94,714 & 521,200 & 18\% \\
		BRAM & 350 & 835 & 42\% \\
		DSP & 840 & 1,680 & 50\% \\ \hline
	\end{tabular}
	\caption{
		Top-level resource utilization of the firmware in an XC7K420T FPGA.
		Two RF-ICE combs at a multiplexing factor of 1024x each fit in a Kintex-7 420T FPGA; one is present in the resource figures shown here.
	    Although present RF-ICE deployments require only a single synthesizer, this design includes a supplemental synthesizer per comb (yellow in \Cref{fig:floorplan-top}) that can be used for dynamic nulling of a HEMT; if it is unnecessary, removing this block would result in additional resource savings.
	}
	\label{tab:resource-usage}
\end{table}

\subsection{Software}
\subsubsection{FPGA interface}

Embedded C++ code running on the IceBoard's ARM CPU forms an essential part of the system, converting requests expressed in an user-friendly software interface to a sequence of register-level interactions with the FPGA.
This interface is also an essential part of the system's maintenance and upgrade path, because it allows the underlying implementation on the FPGA's signal path to change freely without user impact, provided that the C++ code absorbs the differences and exposes a consistent API.

All signal-path controls are exposed via a JSON RPC protocol \cite{Smecher2012}.
This protocol is designed to interface cleanly to Python method-call semantics running on a control PC, but is not tied to any particular client language or runtime.
The underlying runtime on the IceBoard is implemented in C++, but is broadly comparable to the C runtime described in Refs.~\citenum{Smecher2012, Bandura2016}.
A Python client library, described in the following section, allows parallel method-call dispatch across multiple resources on multiple IceBoards.

\subsubsection{Control software}\label{sec:hidfmux}

RF-ICE provides a Python library as the primary user control interface. This high-frequency digital frequency multiplexing (\textit{hidfmux}) control software runs on a computer external to the IceBoard and is designed to interface the digital elements of the ICE system with hardware readout resources. Paralleling the ICE framework's \textit{IceCore} package\cite{Bandura2016}, the library represents physical and digital elements of the readout system as database objects with associated attributes and methods that can be accessed by the user, and enables parallel execution of methods across multiple readout objects. It is designed for control over large MKID arrays, as well as to facilitate smaller-scale testing and characterization of prototype devices.

\paragraph{\textit{hidfmux} hardware map.} Because the primary intent of the software is control over physical and digital elements of the readout system, the initialization of a session in the control software begins with the establishment of a set of Python objects corresponding to resources in the readout chain: a hardware map. This map is a relational database implemented in Structured Query Language (SQL), which links objects in the readout chain based on their physical and logical connections. For example, an IceBoard object is linked to two ReadoutModule objects (corresponding to the dual synthesis/demodulation pathways on the AD9082 mezzanine and RF-ICE firmware), which in turn are each are linked to 1024 ReadoutChannel objects, each of which generates and demodulates a tone and contains its own frequency and amplitude attributes. Commands to collections of hardware map objects can be made asynchronously, allowing efficient parallel control over many resources at once. The use of such a database engine, along with effectively simultaneous control over multiple objects, allows the software to control readout systems at a wide range of complexity and scale without greatly increasing the complexity or latency experienced by the user, from a single IceBoard operating a handful of test detectors to an array of IceBoards operating the SPT-3G+ focal plane.

\begin{figure}
    \centering
    \includegraphics[width=0.7\linewidth]{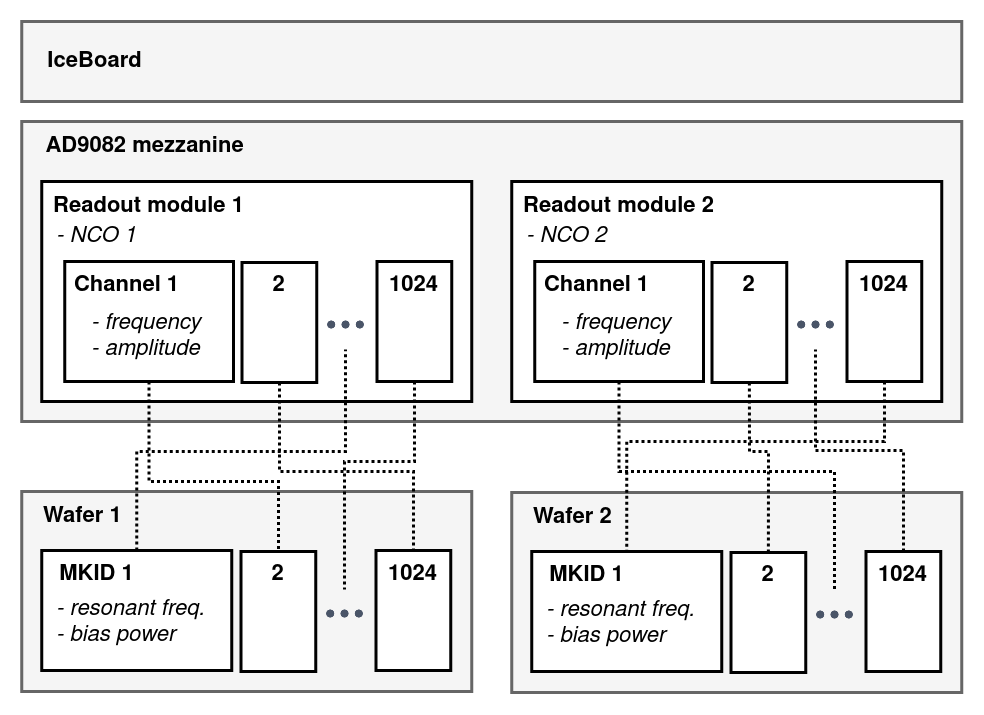}
    \caption{Depiction of \textit{hidfmux} database objects for key resources in the readout chain of a single IceBoard. \textbf{Bold} text names the object, with example key attributes for the object type in \textit{italics}. While in the physical readout system, each module is connected to a detector wafer via a single coaxial line, in the software, individual channels are mapped to MKID objects, each associated with a wafer object, for bookkeeping of detector properties and allowing the efficient initialization of suitable carriers. }
    \label{fig:hardware-map}
\end{figure}

\begin{figure}
    \centering
    \includegraphics[width=\textwidth]{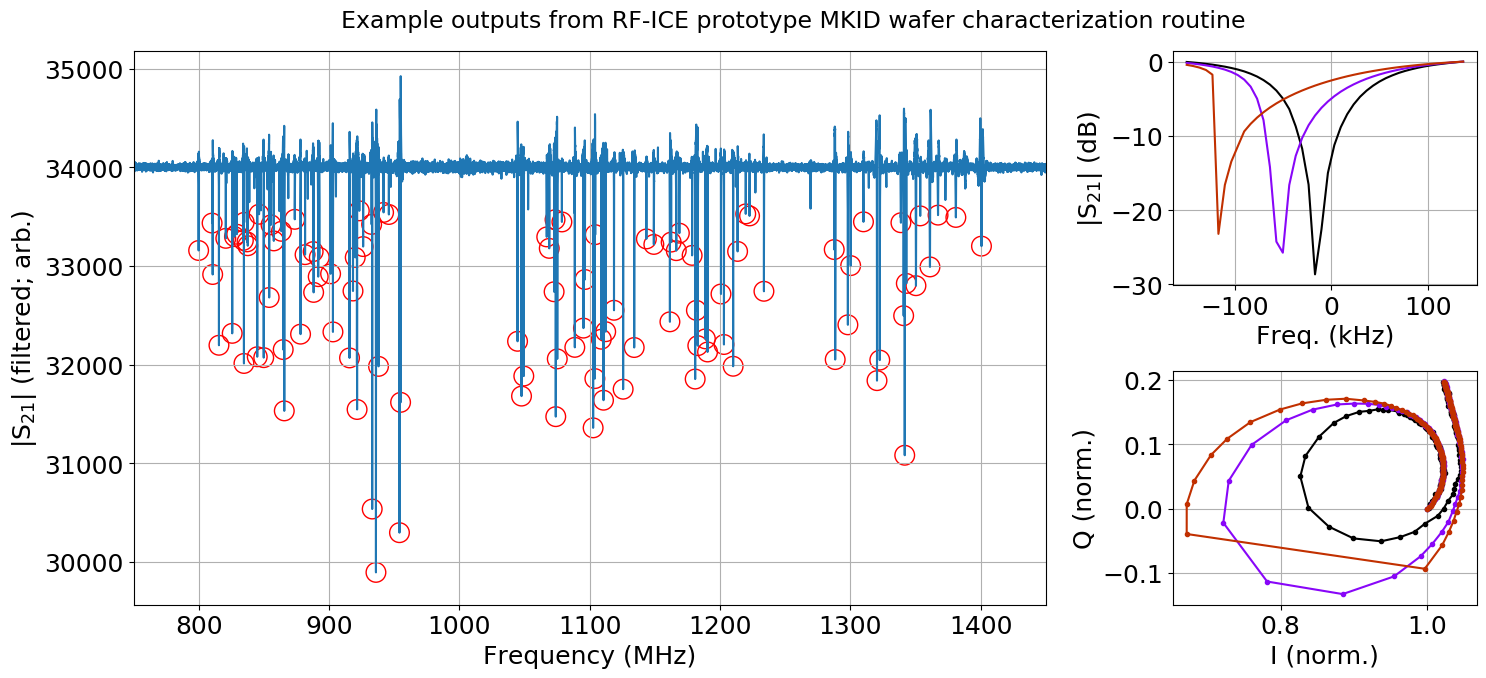}
    \caption{Example data products from the RF-ICE system, in operation characterizing a prototype MKID wafer. The left panel shows an output of the resonance finder, which measures the $S_{21}$ response of the system (blue trace) and uses filtering and a peak finding algorithm to locate resonant features associated with the detectors on the wafer (red circles). Once identified, each resonator's optimal carrier bias power is determined by performing localized frequency sweep measurements across it at a range of probe powers as shown in the right two panels, with warmer colours corresponding to larger carrier powers; top showing the magnitude of the response as a function of frequency, bottom showing the quadrature-demodulated I and Q components of the same data plotted on the I-Q plane. Fitting an appropriate detector model to the data, its optimal bias point is identified by determining the power at which its responsivity is maximized -- this occurs at the maximal rate of change of the trace on the I-Q plane relative to frequency. At carrier powers above this, the resonance exhibits ``bifurcation'' behaviour (for example, yellow trace in right-hand panels), where its state may jump between two stable solutions. This process can be done for multiple detectors simultaneously, allowing rapid characterization of the entire array.}
    \label{fig:data_products}
\end{figure}

A hardware map for a readout system involving a single IceBoard (such as the one whose use in characterizing prototype MKIDs is described in Section \ref{sec:operation}) is depicted in Figure \ref{fig:hardware-map}. Each readout module is associated in software with a detector wafer to which it is physically connected via coaxial line. Software mappings between readout channels and MKID objects allow the bookkeeping of parameters for each detector, such as bias frequency and power, which are determined during a preliminary measurement of the wafer, example data products from which are shown in Figure \ref{fig:data_products}.

Once the parameters are recorded, carrier tones are generated according to the mapping. The state of each detector is read out by monitoring the returning amplitude and phase of the quadrature-demodulated carrier channel associated with it. Under changing loading conditions, the parameters saved in the mapping can be re-measured and updated as required. 

\paragraph{Measurement algorithms.} 

The hardware map and object model provide a versatile foundation on which algorithms of varying complexity can be built. For instance, a user may directly set a single carrier's frequency and amplitude by accessing those attributes in the corresponding ReadoutChannel object, and then collect and save a time-ordered series of samples from that channel. However, it is often useful to perform more complex tasks, such as a two-port network analysis, and the \textit{hidfmux} library provides several algorithms of this type.

The network analysis routine is frequently used when first examining a system under test, such as a new detector wafer. Similar to an $|S_{21}|$ measurement with a traditional vector network analyzer, the ICE system records an amplitude and phase at each frequency, in response to an input signal tone. Unlike a typical traditional network analyzer, however, each module of the ICE system operates 1024 probe tones simultaneously, each measuring a frequency and probe amplitude which can be individually assigned. The user can therefore be creative with their placement; for example, given a set of known frequencies corresponding to the resonance locations of detectors on a wafer, each module in the system can perform simultaneous frequency sweep measurements at up to 1024 separate frequency locations. Performing such a parallel sweep measurement across the resonant features corresponding to a set of MKIDs allows rapid characterization of the entire detector array. Applying a suitable detector model to the data, extracted parameters such as resonant frequency, optimal bias power, Q values, etc, can then be stored in \textit{hidfmux} MKID objects in a saved hardware map, which can be loaded into the program to automatically initialize a future session. Example data products from this measurement routine are shown in Figure \ref{fig:data_products}.

These algorithms can themselves be combined into larger measurement routines, expediting the characterization of prototype MKIDs. Developing alongside the detectors themselves, they will provide the foundations for the future use of the system in controlling large-scale on-sky detector arrays.

\section{In operation with MKID prototypes}\label{sec:operation}

The RF-ICE readout platform is in active development, and has been baselined for deployment on two upcoming South Pole Telescope experiments: SPT-SLIM and SPT-3G+. With target deployment dates of 2023 and 2024/2025, respectively, these two experiments are themselves undergoing active development. The design, testing, and refinement of the RF-ICE readout system has therefore been done in tandem with that of the projects as a whole. This has proved particularly synergetic in the case of the detector design and fabrication, as the RF-ICE system's performance, interface, and capabilities have been both driven and tested by trial use in characterizing detector prototypes.

\begin{figure}
    \centering
    \includegraphics[width=0.4\linewidth]{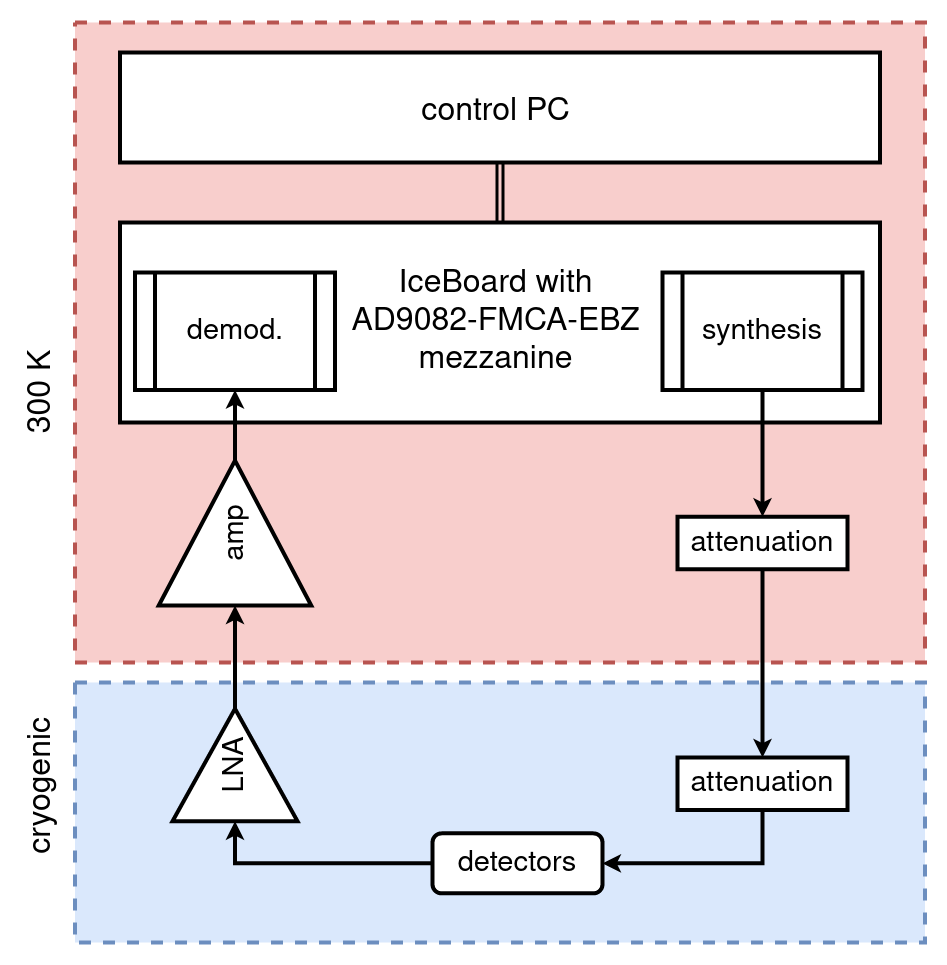}
    \caption{Schematic of the RF-ICE analog signal chain used to interface the digital electronics with a detector wafer under test. The path of the carrier tones is indicated by arrows, passing through several stages of attenuation on the input line to the cryogenic detectors, before being amplified on the return. Elements in the chain are chosen to scale the carriers to suit both the optimal bias power of the detectors, as well as to maximally utilize the dynamic range of the AD9082's ADC and DAC, which are the interfaces to the ICE system's demodulation and synthesis chains, respectively.}
    \label{fig:ext_sig_chain}
\end{figure}

To operate the prototype detectors, the RF-ICE motherboard and mezzanine are paired with a system of external hardware and interfaced with a cryogenic test environment. This external system is representative of what will be used in a deployment context, and so also serves as a test-bed for the final analog electronics configuration. The system, depicted in Figure \ref{fig:ext_sig_chain}, is comparatively simple versus readout electronics for previous generations of SPT cameras, and consists only of a series of attenuators, the detectors themselves, a cryogenic low-noise amplifier (LNA), and room-temperature amplification. The primary function of each electronic element of the chain is to scale the carrier tones to provide optimal bias conditions for the detectors while fully utilizing the dynamic range of the AD9082's ADC and DAC. The simplest case of this testing is done using detectors which are not optically coupled, to avoid the complications of varied optical loads.

This electronics configuration is considered a single readout chain, and is an independent, scalable, and complete unit capable of operating up to 1024 detectors. In a deployment context, multiple readout chains operate together to allow control over the complete focal plane.

In assembling the analog signal chain, two key quantities are of interest: signal level and noise. The readout system must supply carrier tones to the detectors at an optimal power level in order to maximize their responsivity, and contribute a low enough total noise power such that the system remains photon-noise limited. The current design targets an optimal MKID bias power of -100 dBm per resonator.

\paragraph{Carrier levels.}

Evenly distributing the DAC's output power amongst 1024 carriers, each carrier leaves the mezzanine at approximately -40 dBm. Several stages of attenuation, distributed across the temperature stages of the test cryostat, combined with small losses in the coaxial cabling, result in a signal attenuation of approximately -60 dB, in the target frequency range from 500 to 1500 MHz. The resulting carriers arrive at the resonators at -100 dBm per tone. Assuming a typical resonance depth of -10 dB, the modulated carriers leave the detectors at approximately -110 dBm.

Superconducting NbTi coaxial cable transports the carriers to the first stage of amplification, +34 dB provided by a CryoElec\footnote{\url{www.cryoelec.com}} low-noise amplifier \cite{cryoelec} operating on the 4 K temperature stage. After amplification, the modulated carriers exit the cryostat and are further amplified at room-temperature to suit the dynamic range of the ADC. As the signal levels have already been significantly amplified, the choice of component for these warm amplifiers is less stringent. The RF-ICE test setups currently in operation use two Mini-Circuits ZX60-3018G-S+ amplifiers. \cite{minicircuits}
\paragraph{Noise.}

The primary noise sources in the RF-ICE readout chain are the DAC and ADC on board the AD9082 mezzanine, and the amplifiers in the analog chain. Comparing their total noise contribution to the expected photon noise of the detectors serves as a metric of success for this aspect of the readout system.

To make this comparison, we first refer each noise contribution to the input of the LNA (i.e., the output of the detectors); these quantities are shown in Table \ref{tab:noise}.

\begin{table}
  \begin{center}

    \begin{tabular}{l|c|c} 
      \textbf{Source} & \textbf{Noise contribution at source} & \textbf{Referred to LNA input }\\
       & (dBm/Hz) & (dBm/Hz) \\
      \hline
      DAC & -120 & -192\\
      LNA & -192 & -192\\
      Warm amplifier & -174 & -211\\
      ADC & -140 & -218 \\
    \end{tabular}
     \caption{Primary electronic noise sources in the RF-ICE test system, listed both at their source and as referred to the LNA input (immediately following the detectors). DAC noise is given as the single sideband phase noise on a -40 dBm carrier tone, at a frequency of 1 Hz offset from the tone, and the depth of the resonant features associated with the detectors under test is assumed to be approximately -10 dB.}
         \label{tab:noise}
  \end{center}
\end{table}

\begin{figure}
    \centering
    \includegraphics[width=0.6\textwidth]{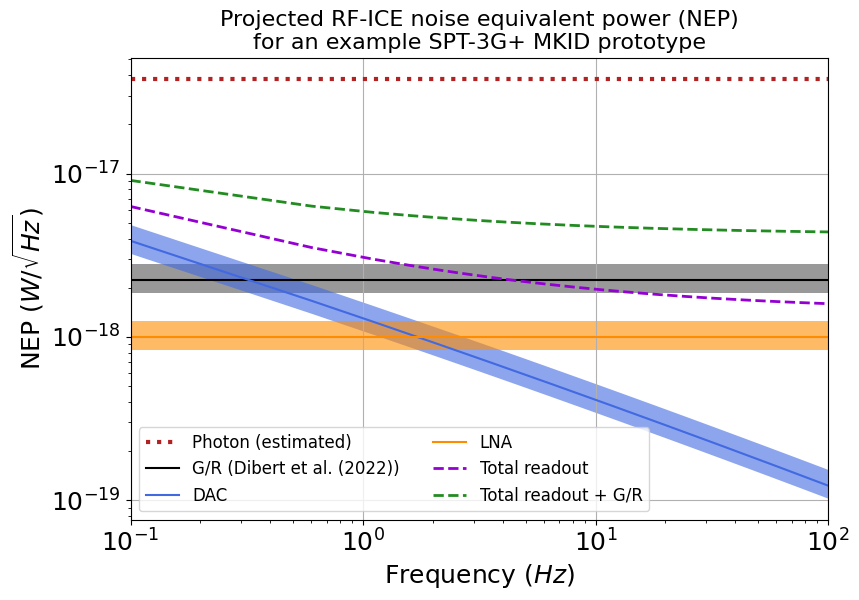}
    \caption{Noise equivalent power (NEP) of key noise sources in the RF-ICE system (as listed in Table \ref{tab:noise}) for an example 220 GHz SPT-3G+ resonator as described in \cite{kariaLTD}, alongside measured generation/recombination (G/R) noise level from the same work. Photon noise is estimated for the same detector under the expected optical power at 220 GHz for the SPT-3G+ instrument of 5 pW. As prototype detector properties will evolve as the experiments move closer to deployment, these values are approximate; an illustrative 20\% uncertainty has been added to the value of the detector responsivity, to demonstrate its impact on system noise projections (indicated by coloured regions). Green and purple dashed lines indicated summed noise quantities, where the values summed are the upper bounds of the shaded regions for each noise contribution. Contributions from the ADC input noise and warm amplification stages are negligible and have been left off for clarity.}
    \label{fig:NEP}
\end{figure}

Noise in an MKID manifests in the normalized extent of its shift away from its unloaded resonant frequency: its fractional frequency shift. This shift can be incurred by any source, whether sky signal, the detector's own quasiparticle generation/recombination (G/R) noise, or noise from other sources in the system. To compare the impacts of noise sources in the readout system, therefore, we first use the detector's frequency response to convert the electrical noise power into a fractional frequency shift in the detector's resonance. This results in a noise spectral density for each quantity, given by \cite{petethesis}:

\begin{equation}
    S_{x,n} = \frac{P_n Q_c^2}{P_c Q_r^4}
\end{equation}

\noindent
where $Q_c$ and $Q_r$ are the coupling and global quality factors, respectively, of the detector's resonant feature, $P_c$ is the power of the carrier tone at the resonator, and $P_n$ is the noise power of the electronic element in question. We can compare these spectral densities with an estimate of the detector's photon noise by converting each to a noise-equivalent power (NEP) using the detector's dark responsivity, as:

\begin{equation}
    NEP_n = \sqrt{S_{x,n}} R_{dark}^{-1}
\end{equation}

where the dark responsivity, $R_{dark}$, as given in Equation 1 of Dibert et al. (2022) \cite{kariaLTD}, is obtained through separate measurement of detector properties and response to changing temperature. Using values from a representative SPT-3G+ detector prototype described in that work for the responsivity, and estimating $Q_c$ and $Q_r$ to be both approximately $10^5$, the estimated NEP values for the noise quantities of Table \ref{tab:noise} are plotted in Figure \ref{fig:NEP}. These are shown alongside, for the same detector, the estimated NEP of the measured quasiparticle generation/recombination noise (the fundamental noise limit of an MKID) and the projected photon noise, also from \cite{kariaLTD}. These estimates do not include a contribution from two-level system noise, which is expected to increase the total noise floor, as it has not yet been characterized for this type of detector.

As detector properties vary under optical loading \cite{petethesis} and the characteristics of the prototypes will evolve as the experiments move closer to deployment, these projections are approximate, but indicate that the system is designed to achieve background-limited noise performance.



\section{Conclusion}

With up to 2048 detector channels per IceBoard, an order of magnitude increase over its predecessor, the RF-ICE system extends the established ICE readout platform into a new operational domain, enabling the high multiplexing factors and gigahertz carrier frequencies required by the upcoming generation of high-density MKID focal planes.

The system is under continuing development, and is currently in use in test environments at Fermi National Accelerator Laboratory, Argonne National Laboratory, the University of Chicago, and McGill University, characterizing prototype detectors for the upcoming SPT-SLIM and SPT-3G+ experiments. As these experiments move closer to commissioning, the existing small-scale RF-ICE test-beds will be expanded and integrated with the deployment hardware, allowing the eventual operation of the 10k-detector SPT-SLIM spectrometer and the planned 35k-detector SPT-3G+ focal plane.

\section{Acknowledgements}
The McGill authors acknowledge funding from the Natural Sciences and Engineering Research Council of Canada and Canadian Institute for Advanced Research.
Work supported by the Fermi National Accelerator Laboratory, managed and operated by Fermi Research Alliance, LLC under Contract No. DE-AC02-07CH11359 with the U.S. Department of Energy. The U.S. Government retains and the publisher, by accepting the article for publication, acknowledges that the U.S. Government retains a non-exclusive, paid-up, irrevocable, world-wide license to publish or reproduce the published form of this manuscript, or allow others to do so, for U.S. Government purposes.

\bibliography{bibliography}
\bibliographystyle{spiebib}

\end{document}